\documentclass[prd,twocolumn,superscriptaddress]{revtex4}
\usepackage{epsfig}
\usepackage{mathrsfs}
\usepackage{amsmath}
\usepackage{amssymb}
\usepackage{color}
\usepackage{graphicx}

\begin{document}

\title {Absorption of Very High Energy Gamma Rays in the Milky Way}

\author{Silvia Vernetto}
\email{vernetto@to.infn.it}
\affiliation{
  Osservatorio Astrofisico di Torino  INAF,
  via P.Giuria 1, 10125 Torino, Italy}

\affiliation{
  INFN, Sezione Torino,
  via P.Giuria 1, 10125 Torino, Italy}

\author{Paolo Lipari}
\email{paolo.lipari@roma1.infn.it}
\affiliation{INFN, Sezione  Roma ``Sapienza'',
  piazzale A.Moro 2,  00185 Roma, Italy}

\begin{abstract}
Galactic gamma ray astronomy at very high energy 
($E_\gamma \gtrsim 30$~TeV) is a vital tool in the study
of the nonthermal universe. 
The interpretation of the observations in this energy region
requires the precise modeling of the 
attenuation of photons due to
pair production interactions ($\gamma \gamma \to e^+ e^-$)
where the targets are the radiation fields
present in interstellar space.
For gamma rays with energy $E_\gamma \gtrsim 300$~TeV 
the attenuation is mostly due to the photons of the 
cosmic microwave background radiation.
At lower energy the most important targets 
are infrared photons with wavelengths 
in the range $\lambda \simeq 50$--500~$\mu$m emitted by dust. 
The evaluation of the attenuation requires a good
knowledge of the density, and energy and angular distributions of the 
target photons for all positions in the Galaxy. 
In this work we discuss a simple model for the infrared
radiation that depends on only few parameters associated to the space
and temperature distributions of the emitting dust.
The model allows to compute with good accuracy 
the effects of absorption for any space and energy distribution 
of the diffuse Galactic gamma ray emission. 
The absorption probability due to the Galactic infrared radiation
is maximum for $E_\gamma \simeq 150$~TeV, and can be as large as
$P_{\rm abs} \simeq 0.45$ for distant sources on lines of sight that pass close
to the Galactic center. The systematic uncertainties on the absorption
probability are estimated as $\Delta P_{\rm abs} \lesssim 0.08$.
\end{abstract}


\maketitle


\section{Introduction}

The past two--three decades have seen the remarkable development 
of high energy gamma ray astronomy with important results obtained
by telescopes, in space and at ground level, 
observing in the energy range from 1~MeV to almost 100~TeV 
\cite{Hinton:2009zz,Funk:2015anl}.
The astrophysical photon fluxes fall rapidly with energy, and
only few photons have been detected with 
$E_\gamma \gtrsim 30$~TeV. On the other hand the exploration of the 
gamma ray sky in the region $E_\gamma \simeq 10^2$--$10^4$~TeV
is a natural and important extension for
high energy astrophysics studies, with clear and strong 
scientific interest.

A new generation of ground--based gamma ray telescopes, 
such as the Cherenkov Telescope Array (CTA) \cite{cta}, 
the extensive air shower array LHAASO \cite{lhaaso} 
and the nonimaging Cherenkov detector HiSCORE \cite{hiscore}
will soon begin to explore the energy range beyond 100~TeV.
These instruments will extend the measured spectra 
of known gamma ray emitters, and have the potential to discover
new classes of sources. Besides the studies of pointlike and
quasi--point--like sources, future detectors with a large field of view
(as LHAASO and HiSCORE) will have the potential
to measure diffuse gamma ray fluxes, such as those 
produced by the interactions of Galactic cosmic rays in 
interstellar space, or generated by an ensemble of unresolved weak sources,
or perhaps, more speculatively, by the decay of Dark Matter particles
in our Galaxy. Of particular interest is the search for 
 gamma ray counterparts to the astrophysical neutrino 
signal recently observed by IceCube in the range $E_\nu \simeq 30$--2000~TeV
\cite{icecube1,icecube2}.

Gamma rays in the energy range we are discussing ($E_\gamma \gtrsim 30$~TeV)
suffer non--negligible absorption during their propagation from the 
emission point to the Earth.
The mechanism that generates the absorption is the creation 
of electron--positron pairs in photon--photon interactions
($\gamma \gamma \to e^+ e^-$), where the target is provided 
by the radiation fields present in interstellar and intergalactic space.
Additional absorption can also occur in the source itself and/or 
in the circum--source environment.

The largest source of target photons is the cosmic microwave background radiation
(CMBR) with a homogeneous number density of approximately 410~cm$^{-3}$.
The extragalactic background light (EBL), generated by the emission of all 
radiation sources in the universe, also fills uniformly space
with an average density of approximately 1.5~cm$^{-3}$ of higher energy photons.
Absorption by the CMBR and EBL essentially precludes extragalactic gamma astronomy
above 30~TeV, on the other hand Galactic astronomy remains possible, but the
interpretation of the observations requires taking into account the effects
of photon absorption.

The main source of target photons for the propagation of gamma rays in the Galaxy
(together with the accurately known CMBR) is the infrared emission of dust, 
heated by stellar light. The radiation field
generated by the dust emission has nontrivial energy,
space and angular distributions 
that must be carefully modeled to compute the absorption effects.
Starlight photons have a smaller number density and their
effects on gamma ray absorption are in good approximation negligible.

The calculation of the interstellar radiation field (ISRF)
is a difficult tasks that requires: (a) the description of the spatial 
distribution of stars (for all spectral types) in the different Galactic structures 
(disk, halo, spiral arms, bar and bulge), (b) the description of the space
distribution and properties (composition, shape and dimensions of the grains)
of interstellar dust, and finally (c) the modeling of the physical processes
(scattering, emission and absorption) that occur in the interactions
between radiation of different wavelength and the dust.

A detailed model of the ISRF, constructed following the points listed above
has been constructed by Strong, Moskalenko and Reimer \cite{strong2000}.
The model was later revised by Porter and Strong \cite{porter2005}, and used
to evaluate the gamma ray absorption in the Galaxy by Zhang et al. \cite{zhang2006}.
More recently Moskalenko, Porter and Strong (hereafter MPS) \cite{moska2006},
have presented a new more precise calculation of the
ISRF and of gamma ray attenuation in the Galaxy. 

The numerical code for the propagation  of relativistic particles in the Milky Way
GALPROP \cite{galprop,Strong:2007nh,Grenier:2015egx}
also implements  the  model of the ISRF  of \cite{porter2005}.
The publicly available version of the GALPROP code has been recently
used by Esmaili and Serpico (hereafter ES) \cite{esma2015}
to obtain the radiation density in different points of the Galaxy,
and the results have been used to evaluate
the absorption of high energy Galactic gamma rays (with
the approximation of an isotropic 
angular distribution of the target photons). The work of Esmaili and Serpico discusses
a particular model for the emission of high energy gamma rays in the Galaxy
(the decay of a very massive particle that forms the galactic matter).

In this paper we perform an independent  evaluation
of the absorption probability   for gamma rays  propagating
in the Galaxy.  The crucial element  in the calculation
is the estimate of the Galactic  dust emission, which is
performed following the work of Misiriotis et al. \cite{misi2006}, who 
have constructed an emission model that depends on only few parameters,
associated to the space and temperature distributions of the interstellar dust. 
This  dust emission model can describe well the main features of the spectral data
of COBE--FIRAS \cite{Wright:1991zz,Reach:1995fy,firas1999}
and the sky maps of COBE--DIRBE \cite{Arendt:1998aj}.
Our approach allows to compute rapidly and with good accuracy
the effects of absorption for any space distribution of
the high energy gamma ray emission.

This work is organized as follows:
in the next section we describe the mechanism of gamma ray absorption,
and discuss the order of magnitude of the effect.
In Sec.~\ref{sec:interstellar}
we outline our model for the interstellar radiation fields in the Galaxy.
In Sec.~\ref{sec:absorption} we use the results to compute the absorption
probability for gamma rays propagating in the Galaxy.
Finally, in the last section we summarize and discuss the accuracy and 
the limits of our calculation.

\section{Attenuation of gamma rays}
\label{sec:attenuation}

\subsection{Pair production cross section}
Given a gamma ray of energy $E_{\gamma}$ and a target 
photon of energy $\varepsilon$, 
the cross section for pair production $\gamma\gamma$ $\rightarrow$ $e^+e^-$
can be expressed as a function of the adimensional 
variable $x=s/(4m_e^2)$, where $s = 2 E_{\gamma} \varepsilon \, (1-\cos\theta)$ 
is the square of the center of mass energy,
$m_e$ is the electron mass, 
and $\theta$ is the angle between the directions of the interacting photons.
The velocity $\beta$ of the $e^\pm$ in the center of mass frame 
is $\beta = \sqrt{1-1/x}$ and the cross section can 
be written as a function of $\beta$ as
\begin{equation}
 \sigma_{\gamma\gamma} = \sigma_{\rm T} \; \frac{3}{16}(1-\beta^2)
 \left[ 2\beta(\beta^2-2)+(3-\beta^4) \, \ln\frac{1+\beta}{1-\beta} \right]
\label{eq:sigma}
\end{equation}
where $\sigma_{\rm T}$ is the Thomson cross section.
\begin{figure}[t]
\includegraphics[width=8cm]{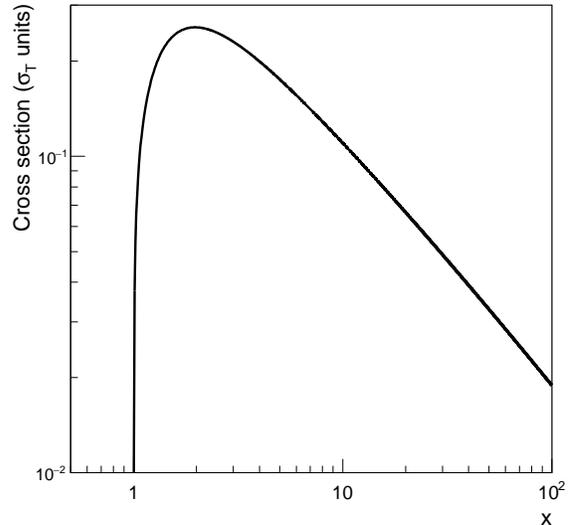}
\caption{Pair production cross section plotted as a function of the variable 
$x = s /(4m_e^2)$. $\sigma_{\rm T}$ is the Thomson cross section.}
\label{fig:crossec}
\end{figure}

Figure~\ref{fig:crossec} shows $\sigma_{\gamma\gamma}$ as a function of 
$x$. The cross section vanishes below the threshold $x =1$ that corresponds
to the center of mass energy $\sqrt{s} = 2 \, m_e$.
Above threshold the cross section increases very rapidly 
reaching a maximum at $x \simeq 1.968$
($\sigma_{\gamma\gamma}^{\rm max} \simeq 0.2554~\sigma_{\rm T}
\simeq 1.70 \times 10^{-25}$~cm$^2$).
Increasing $x$ the cross section decreases monotonically,
and for $x$ large it takes the asymptotic behavior $\propto 1/x$ (or more
exactly $\propto \log(x)/x$).

From the shape of the energy dependence of $\sigma_{\gamma\gamma}$ it follows that 
target photons of a given energy $\varepsilon$
are important for the absorption of gamma rays in a relatively narrow
range of energy around $E_\gamma \approx m_e^2/\varepsilon$.
For a fixed value of $\varepsilon$ the gamma ray energy threshold for pair 
production is 
\begin{equation}
E_{\gamma}^{\rm th} = \frac{2\, m_e^2}{\varepsilon \, (1-\cos\theta)} 
\simeq \frac{0.52}{\varepsilon_{eV} (1-\cos\theta)}~ {\rm TeV} 
\end{equation}
where $\varepsilon_{eV}$ is the target photon energy in electronvolts.
Increasing $E_\gamma$ (for a fixed $\theta$)
the cross section grows rapidly reaching the maximum value 
for $E_{\gamma}^{\rm max} \simeq 1.968~E_{\gamma}^{\rm th}$, it is reduced to
half the maximum value at energy $E_{\gamma} \simeq 8.0~E_{\gamma}^{\rm th}$
and then decreases approximately $\propto E_\gamma^{-1}$.

\begin{figure}[t]
\includegraphics[width=8cm]{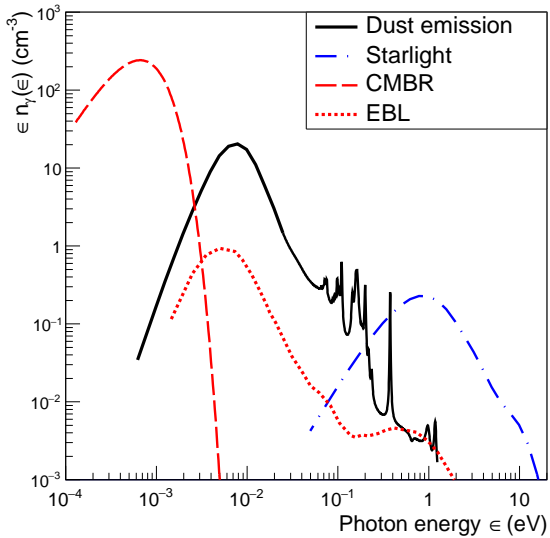}
\caption{Number density   of the different radiation components
  in the solar neighborhood, plotted in the form
  $\varepsilon \; n_\gamma(\varepsilon)$  versus the photon energy
  $\varepsilon$.}
\label{fig:numdens}
\end{figure}
The absorption probability per unit path length (or absorption coefficient)
for a gamma ray of energy $E_\gamma$ and direction $\hat{u}$ 
at the space point $\vec{x}$
can be calculated integrating the interaction probability
over the energy and angular distributions of the target photons:
\begin{eqnarray}
K(E_\gamma, \hat{u}, \vec{x}) 
& = & \int d\varepsilon \int d\Omega ~
 (1-\cos\theta) ~ 
n_{\gamma}(\varepsilon, \Omega,\vec{x}) ~ \nonumber \\
 & ~ & ~~~~~~~~~~~~~ \times \sigma_{\gamma\gamma}[s(E_{\gamma},\varepsilon,\theta)] ~.
\label{eq:kabs}
\end{eqnarray}
In this equation $n_{\gamma}(\varepsilon, \Omega, \vec{x})$ 
is the number density of photons of energy $\varepsilon$ and 
direction $\Omega$ at the point $\vec{x}$, and the
integration is extended to all energies and the entire solid angle. 
The factor $(1-\cos\theta)$ accounts for the relative velocity of the
interacting particles.

\begin{figure}[t]
\includegraphics[width=8cm]{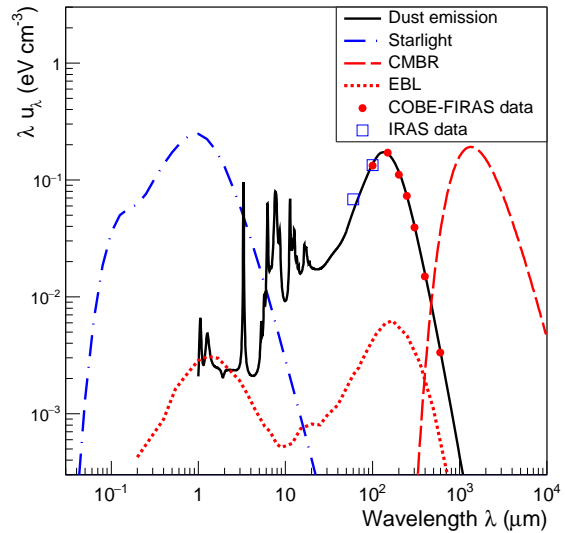}
\caption{Energy density of the radiation fields 
in the solar neighborhood, plotted in the form $u_\lambda \; \lambda$
versus  the wavelength $\lambda$. The points are measurements
by COBE--FIRAS \cite{firas1999} and IRAS \cite{MivilleDeschenes:2004ci}.
}
\label{fig:spectrum_here}
\end{figure}

The survival (absorption) probability for photons of energy $E_\gamma$
traveling between points $\vec{x}_1$ and $\vec{x}_2$ can be written as
\begin{eqnarray}
P_{\rm surv} (E_\gamma, \vec{x}_1, \vec{x}_2) & = &
1- P_{\rm abs} (E_\gamma, \vec{x}_1, \vec{x}_2)
\nonumber \\[0.12 cm]
& = & 
\exp \left [ - \tau(E_\gamma, \vec{x}_1, \vec{x}_2 \right ] 
\label{eq:prob-abs}
\end{eqnarray}
and the optical depth $\tau$ can be calculated integrating the
absorption coefficient $K$ along the trajectory:
\begin{equation}
\tau(E_{\gamma},\vec{x_1},\vec{x_2}) 
= \int_{0}^{|\vec{x}_2 - \vec{x}_1|} ~dt ~ 
K(E_{\gamma}, \hat{u}, \vec{x}_1 + \hat {u} t) 
\label{eq:kint}
\end{equation}
where $\hat{u}$ is the versor parallel to $(\vec{x}_2 - \vec{x}_1)$.
%
For the simple case where the 
target radiation field is homogeneous and isotropic,
the absorption coefficient $K$ is independent from the position 
and direction of the gamma ray, and $\tau$ 
increases linearly with distance:
$\tau (E_\gamma, \vec{x}_1, \vec{x}_2) = K(E_\gamma)~|\vec{x}_2 - \vec{x}_1|$. 

According to equation \eqref{eq:kabs},
the calculation of the survival probability
requires the knowledge of the number density, energy spectrum and
angular distribution of the target photons at all points on the gamma ray
trajectory. 

\subsection{Radiation fields in the solar neighborhood} 
\label{sec:radiation-components}

The radiation field in our Galaxy can be decomposed as the 
sum of four components:
\begin{equation}
n_\gamma = n_\gamma^{\rm CMBR} + n_\gamma^{EBL} + n_\gamma^{\rm stars} + n_\gamma^{\rm dust}
\label{eq:ngamma-decomposition}
\end{equation}
The first two terms in this equation describe 
extragalactic components that permeate uniformly 
the Galaxy and the entire universe and have to very good approximation
an isotropic angular distribution.
The cosmic microwave background radiation 
(CMBR) is the well--known relic of the Big Bang, while
the extragalactic background light (EBL) has been formed by the
emission of all extragalactic sources during the history
of the universe. 
The other two terms in Eq. 
(\ref{eq:ngamma-decomposition}) are of Galactic origin and describe the
radiation emitted by stars and by dust (heated by starlight).
Both Galactic components have nontrivial space and 
angular distributions, the
densities are larger near the Galactic center (GC) and close to the 
Galactic plane, and the photon angular distributions 
trace the shape of the Galactic disk.

The spectrum of the radiation in the solar neighborhood
and the contribution of the different components is shown
in Figs.~\ref{fig:numdens} and \ref{fig:spectrum_here}.
In Fig.~\ref{fig:numdens} the spectrum is shown as
the angle integrated number density of photons $ \varepsilon \; n_\gamma (\varepsilon)$
versus the photon energy $\varepsilon$.
In Fig.~\ref{fig:spectrum_here} the same spectrum is shown in the form
of the energy density $\lambda \; u_\lambda$ versus the wavelength $\lambda$.
Inspecting the figures one can see that three
radiation components (CMBR, dust emission and starlight) have 
spectra with similar shape but maxima 
in different energy (wavelength) ranges and different normalizations.
The fourth component (the EBL) has a shape similar to the
sum of the Galactic star and dust emission, reflecting the fact that
it is formed by the emission of all galaxies in the universe.

The largest component (in terms of number of photons per unit volume) 
of the target radiation field is the CMBR that,
in a very good approximation, is a blackbody spectrum
of temperature $T_{\rm CMBR} = 2.7255$~Kelvin \cite{Fixsen:2009ug}.
This corresponds to a photon density 
of 410.7~cm$^{-3}$ with isotropic angular distribution and 
average photon energy
\begin{equation}
\langle \varepsilon \rangle_{\rm CMBR} =
\frac{\pi^4} {30 \, \zeta(3)} \; k_{\rm B} \, T_{\rm CMBR} 
\simeq 6.34 \times 10^{-4}~{\rm eV} 
\end{equation}
($k_B$ is the Boltzmann constant and $\zeta$ is the Riemann Zeta function).

The second component in order of importance is 
the radiation emitted by Galactic interstellar dust heated by 
stellar light. This emission has been the object of several studies
and reviews \cite{Draine:2003if,draine_li,Compiegne:2010na}.
Most of the energy radiated by dust
is emitted in the far infrared to millimeter range of wavelength
by the largest grains that in good approximation are in
thermal equilibrium with the local radiation field. 
Accordingly the wavelength distribution of the emission 
(for a uniform population of grains) 
is well described by a modified blackbody spectrum:
\begin{equation}
\eta_\lambda \simeq \rho_{\rm dust} \; \kappa_\lambda \; B_\lambda (T) ~.
\label{eq:eta-emission}
\end{equation}
In this equation $\eta_\lambda$ is the power emitted per unit volume, 
unit solid angle and unit wavelength by the dust,
$\rho_{\rm dust}$ is the mass density of the dust, 
\begin{equation} 
B_\lambda(T) = \frac{2 h c^2} {\lambda^5}~\left [{\exp
\left (\frac{hc\lambda}{k_{\rm B} \, T} \right )-1} \right ]^{-1}
\end{equation}
is the Planck function for temperature $T$,
and $\kappa_\lambda$ is the wavelength dependent
dust emissivity cross section per unit mass
that modifies the blackbody shape of the emission.
Empirically the emissivity cross section has often been described 
as a power law:
\begin{equation}
\kappa_\lambda = \kappa_0 \; \left ( \frac{\lambda_0}{\lambda}\right )^\beta ~.
\label{eq:kappa-power}
\end{equation}
Recently the Planck collaboration \cite{planck1} has parametrized its observations
at frequencies $\nu = 353$, 545 and 857~GHz, together with the 100~$\mu$m measurements 
of IRAS as reanalyzed in  \cite{MivilleDeschenes:2004ci} using the simple form 
\begin{equation}
I_\nu = I_0 \; B_\nu (T) \; \left ( \frac{\nu}{\nu_0} \right )^\beta 
\label{eq:infrared-intensity}
\end{equation}
where $I_\nu$ is the energy flux at the frequency $\nu$.
This three--parameter form for the observable energy flux
follows from Eqs. (\ref{eq:eta-emission})
and (\ref{eq:kappa-power}) when absorption is negligible,
as in fact is the case for infrared wavelengths and Galactic distances.
The average of the spectral shape parameters for fits performed for 
all lines of sight in the sky 
are $ \langle T \rangle = 19.7\pm 1.4$~K and 
$\langle \beta \rangle = 1.62\pm 0.10$, where the uncertainties are the dispersions
of the best fit parameters for different sky regions \cite{planck1}.
These results indicate that the energy distribution of the dust emission
(in the range of frequency considered) is well described by a modified backbody 
spectrum, and that the temperature of the emitting dust
has only small variations in different points of the Galaxy.
For the modified blackbody spectrum, the average photon energy
\begin{equation}
\langle \varepsilon \rangle = (3 + \beta) 
\; \frac{\zeta(4 + \beta)}{\zeta(3+\beta)}\; k_{\rm B} \, T \simeq 0.0079 ~\left ( 
\frac{T}{20~{\rm K}}\right ) ~{\rm eV}
\label{eq:eps-mbb}
\end{equation}
scales linearly with $T$, and has a weaker dependence on $\beta$.
The numerical value in the last equality in Eq. (\ref{eq:eps-mbb}) is calculated
for $\beta = 1.62$.

Integrating in frequency and solid angle 
the expression of Eq. (\ref{eq:infrared-intensity}) 
one obtains the energy (number) density
 0.19~eV/cm$^3$ (24.9~cm$^{-3}$). These are in fact 
underestimates of the energy and number density of the radiation emitted by dust, 
because the spectral form of Eq. (\ref{eq:infrared-intensity}) 
is not valid for wavelength shorter than $\lambda \lesssim 50~\mu$m 
(or photon energy $\varepsilon \gtrsim 0.025$~eV).
For short wavelengths, the emission is 
dominated by the contribution of small dust grains, that are not in thermal
equilibrium. This emission depends on the grain size distribution, 
and exhibits prominent lines (clearly visible in Fig.~\ref{fig:numdens}).
We have estimated that the nonthermal contribution to the dust emission
introduces a large correction (approximately 45\%) to the energy density
of the radiation, but a much smaller correction to the number density
of photons of order 5\%--7\%.

The emission of stars in the Galaxy forms an important component 
of the radiation field in the solar neighborhood.
This emission can be described as
the superposition of blackbody spectra 
with a range of temperatures that reflects the mass and age distributions
of stars in Galaxy. 
The energy distribution of starlight shown in Fig.~\ref{fig:numdens} is the 
model of Mathis, Mezger and Panagia \cite{mezger1,Mathis:1983fx} who have 
fitted the (angle integrated) energy distribution of starlight as the sum 
of three diluted blackbody spectra at temperatures of 3000, 4000 and 7500~Kelvin,
estimating the energy density in the solar neighborhood as 0.43~eV/cm$^{-3}$,
with a number density of approximately 0.42 photons/cm$^3$ with average energy
$\langle \varepsilon \rangle \simeq 0.97$~eV. The contribution of young hot stars
emitting in the ultraviolet gives an additional small contribution of
approximately 0.07 photons/cm$^{3}$ with average energy 5.4~eV.

The fourth component of the radiation field is the 
extra galactic background light.
The calculation of the EBL is a difficult task that requires
a description of the injection of radiation from
all galaxies during the entire history of the universe.
Several models exist in the literature
\cite{Franceschini:2008tp,Finke:2009xi,Kneiske:2010pt,Gilmore:2011ks,Stecker:2016fsg},
with results that can differ by a factor as large as $\approx 3$ in the
infrared region.
When gamma rays propagate for extragalactic distances, one has to consider
the cosmological evolution of the EBL and CMBR
(see for example \cite{Batista:2016yrx}).
In the case of the CMBR the
redshift evolution is exactly known, for the EBL it depends on the modeling
of the injection. For propagation in the Galaxy, the
time scale is sufficiently short, so that the redshift
evolution of EBL and CMBR can be safely neglected.

The curve shown in Fig.~\ref{fig:numdens} shows the EBL
model of Franceschini et al. \cite{Franceschini:2008tp} calculated at
redshift $z=0$. 
The EBL angle integrated number density is small, of order 1.5~cm$^{-3}$.
This has the consequence that this component, while very important 
for extragalactic gamma rays, has a negligible effect for the propagation
of photons inside our Galaxy.

\subsection{Gamma ray absorption coefficient in the solar neighborhood}
The absorption coefficient 
for the solar neighborhood can be calculated from Eq.
(\ref{eq:kabs}) using the energy spectra of the target photons 
described above and shown in Fig.~\ref{fig:numdens}, and 
a model for the angular distribution of these photons.
Using the decomposition of the target photon field 
of Eq. (\ref{eq:ngamma-decomposition}) the
absorption coefficient can be written as the sum
of the contributions of the different components:
\begin{eqnarray}
K (E_\gamma, \hat{u}, \vec{x}) & = & 
K_{\rm CMBR} (E_\gamma) + K_{\rm EBL} (E_\gamma) \nonumber \\[0.15 cm]
& + & K_{\rm dust} (E_\gamma, \hat{u}, \vec{x}) + K_{\rm stars} (E_\gamma, \hat{u}, \vec{x}) ~.~~~
\end{eqnarray}
The components associated to the CMBR and the EBL are 
independent from $\vec{x}$ and from the photon direction $\hat{u}$.
To estimate the angular distribution of the radiation emitted by dust
we have used the IRAS map of the 100~$\mu$m radiation given in 
\cite{MivilleDeschenes:2004ci} and made the assumption 
that the distribution in energy and angle of the photons factorizes 
in the form $n_\gamma (\varepsilon, \Omega) \simeq n_\gamma (\varepsilon) \; F(\Omega)$,
where the angular distribution $F(\Omega)$ 
is independent from $\varepsilon$ and equal to the 
(normalized to unity) distribution at 100~$\mu$m.
Comparing the angular distributions of the radiation measured 
at different wavelengths \cite{planck1,MivilleDeschenes:2004ci}
one can see that the factorization hypothesis is a very good approximation.

\begin{figure}[t]
\includegraphics[width=8cm]{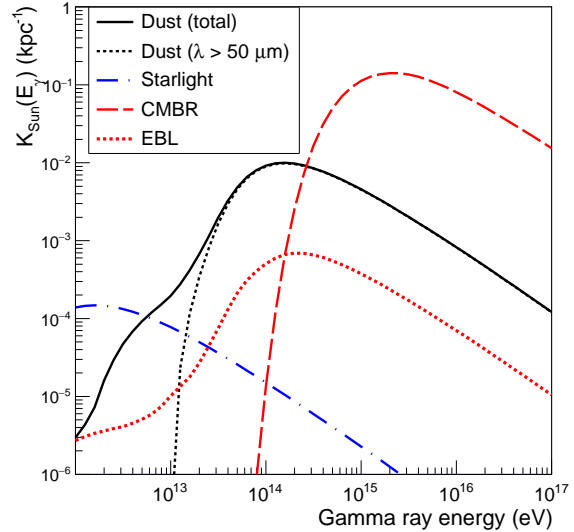}
\caption{Absorption coefficient
 in the solar neighborhood
 (averaged over the photon direction) 
 for the different components of the radiation field. }
\label{fig:kabs-sun}
\end{figure}

The absorption coefficient in the solar neighborhood (averaged over direction)
is shown in Fig.~\ref{fig:kabs-sun}.
It is instructive to briefly discuss the results. 
The contribution of the CMBR has a maximum for a gamma ray energy 
$E_{\rm CMBR}^* \simeq 2.2$~PeV, where the absorption coefficient has the value
$K_{\rm CMBR}^{\rm max} \simeq 0.141$~kpc$^{-1}$. 
These numerical results are easy to understand qualitatively. 
The energy of maximum absorption $E_{\rm CMBR}^*$ is approximately equal 
(it is larger by a factor 1.34) to 
 $4 \,m_e^2/\langle \varepsilon \rangle_{\rm CMBR}$
where $\langle \varepsilon \rangle_{\rm CMBR}$ the average energy of the CMBR photons,
and corresponds to the gamma ray energy that is best matched to the distribution
of the target photons to maximize the pair production cross section.
The maximum value of the absorption coefficient 
$K_{\rm CMBR}^{\rm max}$ is approximately equal to (a factor 0.76 of difference to)
to the quantity 
$\sigma_{\rm T} \; N_\gamma^{\rm CMBR}/4$, which is the product of the
pair production at maximum 
($\sigma_{\gamma\gamma}^{\rm max} \simeq \sigma_{\rm T}/4$)
and the total number density of CMBR photons.
For $E < E^*_{\rm CMBR}$ the absorption coefficient falls very rapidly
reflecting the spectral shape of the CMBR photon distribution, while
for $E_\gamma \gg E^*_{\rm CMBR}$ the absorption coefficient falls 
 $\propto E_\gamma^{-1}$ reflecting the $s^{-1}$ dependence of the pair
production cross section at large $s$.

This pattern for the energy dependence of the absorption coefficient 
is repeated for the components associated to the target radiation 
generated by dust and star emission. 
In these three cases (CMBR, dust emission and star emission) the energy spectrum
of the radiations has (exactly in the case of the CMBR, and approximately 
in the other two cases) a blackbody shape, characterized by temperatures 
of approximately 2.7~Kelvin for the CMBR, 20~Kelvin for the
dust emission and 3000--7500~Kelvin for starlight.
Accordingly, the average energies of the photons associated
to the three components are significantly different: 
($\langle \varepsilon \rangle \simeq 6.3 \times 10^{-4}$, $7.6 \times 10^{-3}$ 
and 0.97~eV for CMBR, dust and star emission, respectively).
This implies that the critical energy $E^*$ where the absorption effects 
are largest is 2.2~PeV for the CMBR, 150~TeV for the dust emission 
and 1.6~TeV for the star emission.
The three components contribute approximately the same energy density
to the target radiation field in the solar neighborhood
($\rho_\gamma \simeq 0.260$, 0.25 and 0.42~eV/cm$^3$), 
(see Fig.~\ref{fig:spectrum_here})
however since the average photon energies are different, 
the (energy and angle integrated) number densities
for the three components are also significantly different 
($N_\gamma \simeq 410.7$, 25.0 and 0.43~cm$^{-3}$).
The maximum absorption coefficient is proportional to the total number density
and is therefore very different for the three components 
($K^{\rm max}_j \simeq 0.141$, $0.98\times 10^{-2}$ and $1.5 \times 10^{-4}$~kpc$^{-1}$).

\begin{figure}[t]
\begin{center}
\includegraphics[width=8cm]{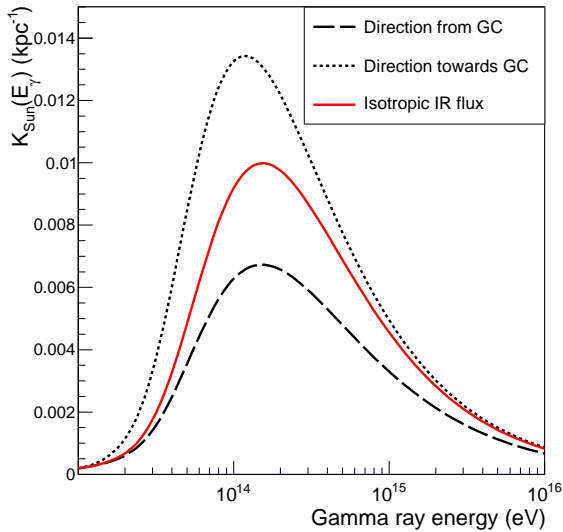}
\end{center}
\caption{Absorption coefficient in the solar
neighborhood due to the dust emitted radiation.
The solid curve is calculated assuming an isotropic angular distribution
of the photons. The other lines are calculated assuming the 
angular distribution observed at 100~$\mu$m by the IRAS satellite,
and for photons traveling radially from
and towards the Galactic center.}
\label{fig:kabs-angular1}
\end{figure}

For an estimate of the importance of the angular distribution of 
the target radiation one can inspect Fig.~\ref{fig:kabs-angular1} that 
shows the absorption coefficient calculated
at the position of the Sun for the lines of sight 
that point towards the center and anticenter of the Galaxy.
The figure also shows for comparison
the absorption coefficient calculated assuming an isotropic
angular distribution for the photons. 
For the line of sight towards the Galactic center
(anticenter) most of the target photons are parallel
(antiparallel) to the gamma ray direction,
and this gives a lower (higher) absorption, with deviations
of order $\mp 30$\% with respect to the result for an isotropic distribution.

As discussed above, the EBL has in first approximation
the same energy distribution of the 
sum of the Galactic dust and star emission components,
but it has a smaller density, reduced by a factor of approximately 20 with respect
to the Galactic radiation field in the vicinity of the Sun. 
It follows that $K_{\rm EBL} (E_\gamma)$ (for the near universe at $z \simeq 0$)
is approximately equal to the sum of the Galactic 
$K_{\rm dust} + K_{\rm star}$ reduced by an order of magnitude. 
This results in a negligible absorption effect for
gamma rays that propagate for Galactic distances.

\subsection{Order of magnitude of gamma attenuation effects}
\label{sec:order-of-magnitude} 
The calculation of the gamma ray attenuation requires a numerical
integration of the absorption coefficient along the gamma ray trajectory.
It is however possible and instructive to obtain a first order estimate
of the size of the effect with some simple considerations.

The order of magnitude of the optical depth due to starlight
along a line of sight of length $L$ can be estimated as:
\begin{equation}
 \tau \approx
 K_\odot ~ L_{\rm eff} (L, \Omega)
 \approx
K_\odot ~ L \; 
\frac{\langle N_\gamma^{\rm star} \rangle }
{N_{\gamma, \odot}^{\rm star}} ~.
\label{eq:tau-stars}
\end{equation}
In this equation
$K_\odot$ is the absorption coefficient at the sun,
$\langle N_\gamma^{\rm star} \rangle$ is the 
density of starlight photons averaged for all points 
along the line of sight and 
$N_{\gamma, \odot}^{\rm star}$ is density in the solar neighborhood.
In Eq. (\ref{eq:tau-stars} the optical depth is proportional
to the absorption coefficient at the Sun position, and to
an effective length $L_{\rm eff}$ that takes into account
the average radiation density along the trajectory.

The maximum absorption coefficient 
due to starlight in the solar neighborhood is of order 
 $1.5 \times 10^{-4}$~kpc$^{-1}$. 
Since the linear size of the Galaxy
is of order 10--30~kpc, even if the density of starlight near the 
Galactic center is 1 order of magnitude larger than
what we observe near the Sun, one can robustly conclude that 
the optical depth for any Galactic source 
is at most of order a few $\times 10^{-2}$, with a small attenuation effect.

An estimate of the optical depth due to the radiation created by
dust can be obtained from the analogous of Eq. (\ref{eq:tau-stars}). 
In this case the maximum of the absorption coefficient 
in the solar neighborhood (obtained for $E_\gamma \simeq 150$~TeV)
is of order 0.0098~kpc$^{-1}$. The calculation of the density of
infrared radiation in different points in the Galaxy is very much 
simplified by the fact that the Galaxy is transparent to infrared radiation,
and it is therefore relatively straightforward
(this will be discussed more carefully in the following) to determine the 
space distribution of the emission from the angular distribution 
of the radiation observed at the Earth, and therefore the density and
angular distribution at any point in the Galaxy. 
It is also straightforward to see that the quantity
$L_{\rm eff} (L,\Omega) 
= \langle N_\gamma^{\rm dust} \rangle/{N_{\gamma, \odot}^{\rm dust}} \; L$ 
converges to a finite result (that depends on the line of sight)
in the limit $L \to \infty$ (because the density of photons vanishes
sufficiently rapidly at large distances from the Galactic center).
In fact most of the gamma ray absorption happens near the GC. 
For lines of sight that pass near the GC the
value of $L_{\rm eff} (\Omega)$ is of order 40--50~kpc, 
while the effective distance to the Galactic center is of order 20~kpc.
The absorption due to the dust emitted radiation corresponds therefore
(for an energy of order 100--150~TeV, where the effect is most important)
to a maximum optical depth of order 0.4--0.5 for objects in
the halo of the Galaxy seen via a line of sight 
that crosses the GC, and to an optical depth of 
order 0.2 for the GC itself. 
A more accurate modeling of gamma ray attenuation due 
to the radiation emitted by dust will be discussed in the following.

The discussion of absorption due to the CMBR 
is straightforward. The maximum absorption coefficient 
(for $E_\gamma^* \simeq 2.2$~PeV) is 0.1415~kpc$^{-1}$, and
optical depth equal to unity corresponds to the distance $L \simeq 7.07$~kpc.
This implies that PeV gamma astronomy is only possible in a sphere
that encompasses a fraction of our Galaxy.
At the energy of 2.2~PeV, photons emitted from the Galactic center
arrive at the Earth with approximately 0.30 survival probability.

\vspace{0.05 cm}
The absorption of each high energy gamma ray generates one electron positron pair
with the constraint that the sum $E_{e^+} + E_{e^-}$ is to a very good
approximation equal to the initial $E_\gamma$. The high energy $e^\pm$ created in
the interaction rapidly lose energy via synchrotron radiation
(because of the Galactic magnetic field) and inverse Compton
scattering (where the target is again the ISRF). Photons created
by these processes have in good approximation no angular correlation
with the initial direction of the gamma ray. As an illustration,
an $e^\pm$ with an energy of 100~TeV in a magnetic field of 3~$\mu$Gauss
(a typical value for the Galactic disk) has a Larmor radius
of order 0.036~pc, while the energy loss length (the distance for which the particle loses
half its energy) is of order of 1~kpc, more than 4 orders of magnitude longer.
The radiation of the $e^\pm$ created by gamma ray absorption
in the Galaxy is part of the Milky Way diffuse emission, but their contribution
is very likely negligible.

\begin{figure}[t]
\includegraphics[width=8.4cm]{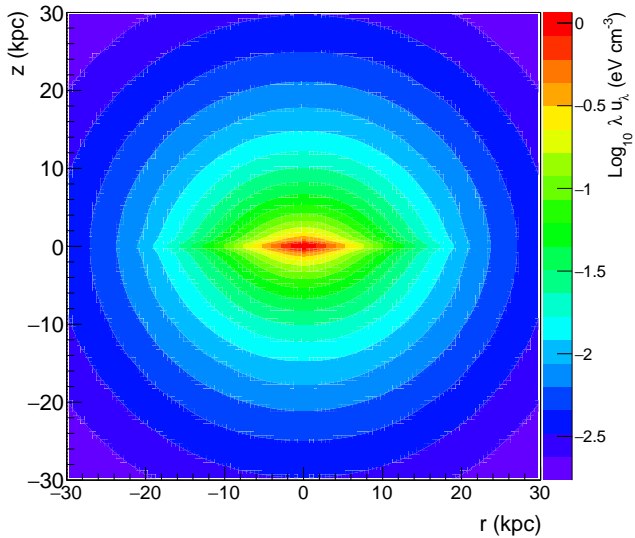}
\caption{Space dependence of the angle integrated energy density
 ($u_\lambda \lambda$) of the infrared radiation
 for wavelength $\lambda$ = 100 $\mu$m, shown as a function of
 the cylindrical coordinates $r$ and $z$
in the Galaxy system.}
\label{fig:IR_zr}
\end{figure}

\section{Interstellar Radiation Fields in the Galaxy}
\label{sec:interstellar}

\subsection{Model for infrared emission}
Since the absorption probability 
of infrared photons traveling Galactic distances is negligible,
the flux $I_\lambda (\Omega, \vec{x})$ that reaches
the point $\vec{x}$ from the direction $\Omega$
can be calculated simply integrating the emission along the line of sight.
Assuming isotropic emission, one has:
\begin{equation}
I_\lambda (\Omega, \vec{x}) = \int_0^\infty dt ~ \eta_\lambda (\vec{x} + \hat{\Omega} \; t)
\label{eq:flux-infra}
\end{equation}
with $\hat{\Omega}$ the versor in the direction $\Omega$, 
and $\eta_\lambda (\vec{y})$ the emission power density
per unit wavelength at the point $\vec{y}$.

A model for the far--infrared emission for interstellar dust in the
Galaxy, which is based on only a small
number of parameters, has been developed by Misiriotis et al.\cite{misi2006},
fitting the spectra and angular distributions reported by
COBE--FIRAS \cite{Wright:1991zz,Reach:1995fy,firas1999} and
COBE--DIRBE \cite{Arendt:1998aj} in the 60--1000~$\mu$m wavelength range.

\begin{figure}[t]
\includegraphics[width=8cm]{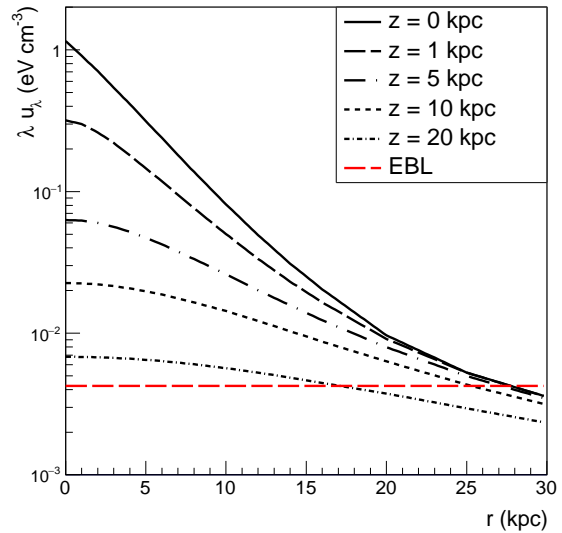}
\caption{Energy density (times wavelength) of the infrared radiation
for $\lambda$ = 100 $\mu$m as a function of $r$ for different values of $z$.}
\label{fig:IR_r}
\end{figure}

In the model the interstellar dust is described as the sum of two components:
``cold'' and ``warm'' dust, that both have a spatial distribution parametrized
by the functional form:
\begin{equation}
 \rho_{c,w}(r,z)= \rho_{c,w}^\circ \exp \left(\frac{-r}{R_{c,w}}
 - \frac{\vert z \vert}{Z_{c,w}}\right) ~.
\end{equation}
In this equation $\rho_c$ ($\rho_w$) is the mass density of the cold (warm) dust,
$r$ and $z$ are cylindrical coordinates in the Galaxy, the 
pairs of parameters $\{R_c, Z_c\}$ and $\{R_w,Z_w\}$ are the exponential
scale height and scale length
of the dust distributions, and $\rho_{c,w}^\circ$ is the dust mass density 
at the Galactic center.
The total mass of the Galactic interstellar dust is estimated as
 $M_{\rm dust}^c \simeq 7.0 \times 10^7~M_\odot$, with the warm dust constituting only
a fraction of order 0.31\%. 

The dust is assumed to be in local thermal equilibrium, 
and therefore to have a well-defined temperature. 
For the cold dust, heated by the diffuse Galactic starlight,
the temperature has an exponential space dependence:
\begin{equation}
T_c(r,z)= (T_c^\circ -T_{\infty}) \exp \left(\frac{-r}{R_T} 
- \frac{\vert z \vert}{Z_T}\right) + T_\infty
\end{equation}
where $T_c^\circ$ (with a best fit value of 19.2~Kelvin) is the temperature at the Galaxy Center, 
 $T_{\infty} = T_{\rm CMBR}$ 
is the temperature at infinity, and $R_T$ and $Z_T$ are parameters that 
give the scale length and scale height of the distribution.

The warm dust is present in star forming regions, and most of the energy
it absorbs is generated by near, young, hot stars. This results
in a constant temperature $T_w \simeq 35$~K \cite{pope2000}.

The dust emission for wavelength $\lambda \ge 50~\mu$m is described
as a modified blackbody spectrum according to Eq. (\ref{eq:eta-emission})
summing the contributions of cold and warm components:
\begin{equation}
\eta_\lambda (\vec{x}) = 
\eta_\lambda^c (\vec{x}) +
\eta_\lambda^w (\vec{x}) ~.
\end{equation}
To describe the dust emissivity $\kappa_\lambda$, 
following Misiriotis et at. \cite{misi2006}
we have used the results given
by Weingartner and Draine \cite{wein2001}, assuming for the dust 
a constant parameter $R_V = 3.1$ (with $R_V$ the ratio of visual extinction 
to reddening for visible light) \footnote{Tables of
numerical values for $\kappa_\lambda$ 
as a function of wavelength are available at https://www.astro.princeton.edu/
\~{}draine/dust/dustmix.html}.

\begin{table}[b]
 \centering
 \vspace*{0.5cm}
 \caption{Best fit parameters to describe the space and temperature
 distribution of interstellar dust in the Milky Way \cite{misi2006}.}
\begin{tabular}{ccc}
\hline
Parameter & Value & Units
\\ 
\hline 
$\rho_c^\circ$ & 1.51 $\times$ 10$^{-25}$ & g~cm$^{-3}$ \\ 
$\rho_w^\circ$ & 1.22 $\times$ 10$^{-27}$ & g~cm$^{-3}$ \\ 
$R_c$ & 5 & kpc \\
$R_w$ & 3.3 & kpc \\
$Z_c$ & 0.1 & kpc \\
$Z_w$ & 0.09 & kpc \\
$T_c^\circ$ & 19.2 & K \\
$R_T$ & 48 & kpc \\ 
$Z_T$ & 500 & kpc \\
$T_w$ & 35.0 & K \\
\hline
\end{tabular}
\label{tab_para}
\end{table}

The model of dust emission
described above is therefore defined by a total of ten parameters:
$\rho_{c,w}^\circ$, $R_{c,w}$, $Z_{c,w}$,
$T_c^\circ$, $R_T$, $Z_T$ and $T_w$. The set of best fit parameters 
obtained by Misiriotis et al. \cite{misi2006}
is reported in Tab.\ref{tab_para}, and 
are those also used by us to describe the radiation.

The dust emission at shorter wavelength (1--50~$\mu$m)
is dominated by the nonthermal contribution of small grains, and
the spectrum is characterized by strong emission lines.
To extend the model to this wavelength region
we have used the work of Draine and Li \cite{draine_li} 
who have calculated the dust emission
for different compositions of the dust grains
and for different models (in intensity and spectral shape)
of the starlight radiation that heats the dust. 
We have made the simplifying assumption that the shape of the 
dust emission in the short wavelength region is equal for all
points in the Galaxy, and described by a single 
Draine and Li model, the one where the
polycyclic aromatic hydrocarbons (the main source of the line emission)
account for 0.47\% of the dust mass density,
and where the dust is heated by a starlight with the 
same spectrum and intensity observed in the solar neighborhood \footnote{
This corresponds to the model \\
{\tt U1.00/U1.00\_1.00\_MW3.1\_00} in the tabulated results made available
in \cite{draine_li}. }.

The intensity of infrared radiation for any direction and
position in the Galaxy can be calculated from the emission model
making use of Eq. (\ref{eq:flux-infra}). 
The energy density of the radiation at a certain point
can be obtained integrating the flux over all directions:
\begin{eqnarray}
u_{\lambda}(\vec{x}) & = & \frac{1}{c} \; \int d\Omega ~I_\lambda(\Omega,\vec{x}) 
\nonumber \\[0.15 cm]
& = & \frac{1}{c} \; \int d^3 y ~\frac{\eta_\lambda(\vec{y}) }{|\vec{x} - \vec{y}|^2} ~.
\end{eqnarray}
The density of infrared photons as a function of the energy $\varepsilon$ can be
obtained transforming these distributions using the
relation $\varepsilon = h \, c/\lambda$. 
The model described here has cylindrical symmetry for rotations around an axis
that passes through the Galactic center, and up--down symmetry
for reflections of the $z$ axis,
so that the density of the radiation depends only on $r$ and $|z|$; and the angular
distributions of the radiation at points (such as the solar system) 
that are on the ($z=0$) Galactic plane have the properties
\begin{equation}
n_\gamma (\varepsilon; b, \ell) =
n_\gamma (\varepsilon; \pm b, \pm \ell)
\end{equation}
with $b$ ($\ell$) the Galactic latitude (longitude) angle.

The space dependence of the energy density of infrared radiation is illustrated in
Figs.~\ref{fig:IR_zr} and \ref{fig:IR_r}.
Figure~\ref{fig:IR_zr} shows the energy density $\lambda \; u_{\lambda}$ 
as a function of the Galactic cylindrical coordinates $r$ and $z$, for the
wavelength $\lambda = 100~\mu$m
(that is close to the maximum of the infrared spectrum) while
Fig.~\ref{fig:IR_r} shows the $r$ dependence of the energy density 
(for the same wavelength) for different fixed values of $z$. 
Most of the infrared flux is concentrated in the region $r \lesssim 10$~kpc
and $|z| \lesssim 2$~kpc.

In this model the space and energy dependence of the radiation field
are approximately factorized, and therefore Figs.~\ref{fig:IR_zr} and \ref{fig:IR_r}
describe well the shape of the space distribution of
the dust emission radiation in a broad range of wavelength.
\begin{figure}[t]
\includegraphics[width=8cm]{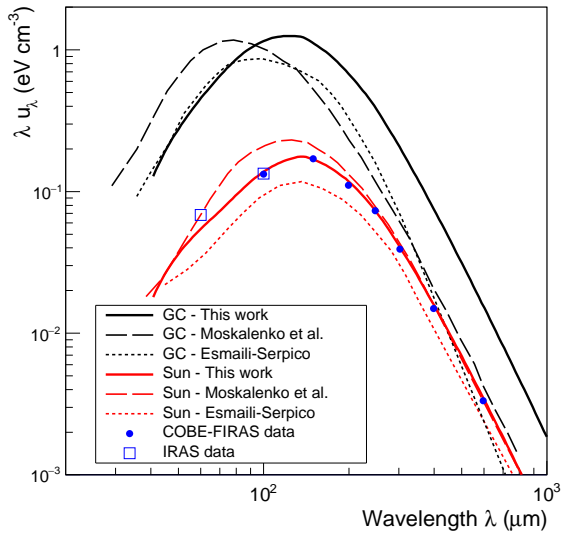}
\caption{
Angle integrated spectra of the dust emitted radiation 
in the vicinity of the Sun and at the Galactic center.
The spectra are shown in the form $\lambda \; u_\lambda$ plotted as
a function of the wavelength $\lambda$.
The spectrum in the solar neighborhood is compared
to the observations of COBE--FIRAS \cite{firas1999} 
(at $\lambda = 100$, 150, 200, 250, 300, 400 and 600~$\mu$m)
and IRAS \cite{MivilleDeschenes:2004ci} 
(for $\lambda = 60$ and 100~$\mu$m). }
\label{fig:spe_fir}
\end{figure}

\begin{figure}[t]
\includegraphics[width=8cm]{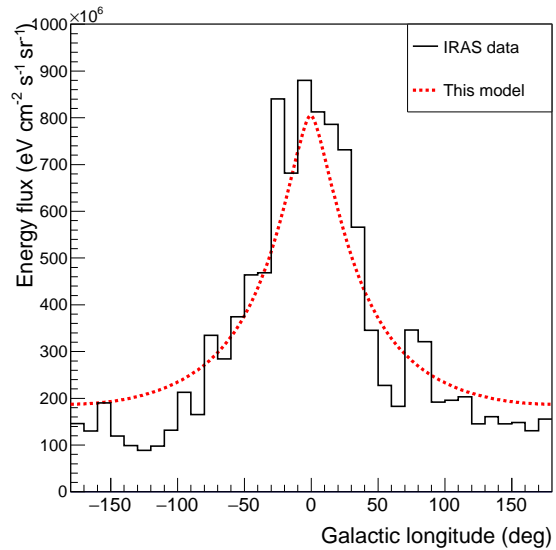}
\caption{Energy flux of the infrared radiation at $\lambda = 100~\mu$m at the Sun position
as a function of the Galactic longitude, compared with the IRAS data
(averaged over 10$^{\circ}$ in longitude).} 
\label{fig:IR_longdis}
\end{figure}

\subsection{Comparison of model with data}

The (angle integrated) energy spectrum of the dust emitted radiation 
in the vicinity of the Sun is shown in Fig.~\ref{fig:spe_fir},
where it is plotted as a function of wavelength and compared to 
the measurements of COBE/FIRAS \cite{Reach:1995fy} 
and IRAS \cite{MivilleDeschenes:2004ci} 
that cover well the wavelength range where 
the radiation is brightest.
The agreement between model and data is very good.
In the figure we also show the radiation spectrum in the solar neighborhood
calculated by MPS \cite{moska2006} and the one obtained using the GALPROP code
by ES \cite{esma2015}. 
These results have a spectral shape similar to the one
calculated in our model, but with normalizations that differ by 
approximately $\pm 30$\% (in opposite directions for the two cases).

Figure~\ref{fig:spe_fir} also shows the infrared spectrum 
at the Galactic center.
In the model discussed here the spectral shape of the radiation
is only weakly dependent on the position in the Galaxy.

\begin{figure}[t]
\includegraphics[width=8cm]{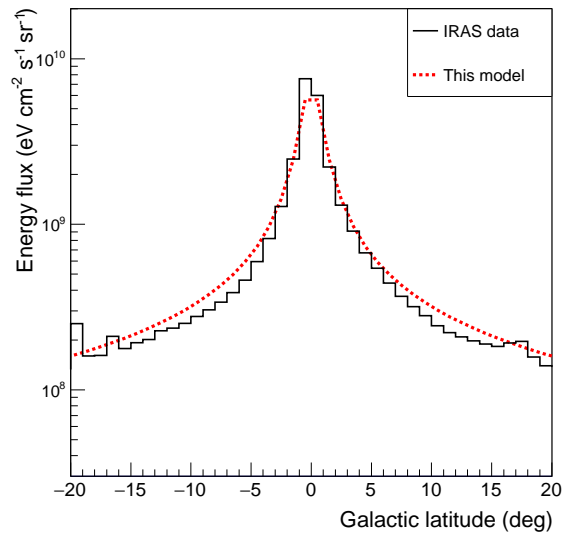}
\caption{Energy flux of the infrared radiation at $\lambda = 100~\mu$m 
at the Sun position
as a function of the Galactic latitude, compared with the IRAS data
(averaged over 1$^{\circ}$ in latitude).}
\label{fig:IR_latdis}
\end{figure}

In Sec.~\ref{sec:radiation-components}
we have noted that the Planck collaboration has
observed that the spectra measured
in the wavelength region between 100 and 850~$\mu$m
along different lines of sights have shapes that are
approximately independent from the direction.
This result is consistent with what we obtain 
in our model, and it is the consequence of the fact that most of
the interstellar dust in the Galaxy has only a narrow range of temperatures
(around 20~Kelvin), with
the warm dust, which is only 0.31\% of the total dust mass,
giving a measurable contribution only at shorter wavelengths.

In the approximation where
the dust has a single temperature in the entire Galaxy
(and therefore the emission spectrum is independent from position),
and using the fact that the absorption of infrared radiation is negligible, 
one has that the spectrum of the dust emitted radiation in different 
points of the Galaxy has a factorized form. The factorization
can be written as
\begin{equation}
n_\gamma (\varepsilon, \Omega, \vec{x}) \simeq n_\gamma^\odot (\varepsilon) ~ F (\Omega, \vec{x}) 
\label {eq:factorization}
\end{equation}
where (without loss of generality) one can take 
$n_\gamma^\odot (\varepsilon)$ as the angle integrated spectrum at the Sun position,
and $F (\Omega, \vec{x})$ is independent from the photon energy (or wavelength).
In the model discussed here the factorization of Eq. (\ref{eq:factorization})
is not exactly valid. The (angle integrated) spectra measured near the
Galactic center are a little harder
than the spectra measured at large distance from the center,
and similarly for a fixed point in the Galaxy,
the spectra in the direction of the GC are a little harder than 
the spectra in the opposite direction;
however the factorization remains a reasonable first approximation
that can be useful as a guide to estimate the absorption effects
(as in our discussion in Sec.~\ref{sec:order-of-magnitude}).

An example of the angular distribution of the dust emitted radiation in our model 
is given in Figs.~\ref{fig:IR_longdis} and \ref{fig:IR_latdis} that show the
Galactic latitude and Galactic longitude distributions in the solar
neighborhood for the wavelength $\lambda = 100~\mu$m. 
The distributions are compared to the map obtained by IRAS \cite{MivilleDeschenes:2004ci}
for the same wavelength.
The infrared photons have a very anisotropic distribution,
with most flux arriving in a narrow range of latitude
around the Galactic equator (approximately 50\% of the flux is contained
in the region $|b| \lesssim 2^\circ$), there is also a significant asymmetry
in longitude, with the flux from the direction of the Galactic center 
approximately 4 times larger than the flux in the opposite direction.
The comparison of the model with the observations shows that the main features of the data 
(the width of the latitude distribution, and the asymmetry between the directions toward the
Galactic center and anticenter) are reasonably well described, but 
on smaller angular scales the model cannot describe accurately
the structures associated to the irregular distributions of stars and dust.

\subsection{Starlight}
To complete the description of the ISRF in the Galaxy it is necessary to 
have a model also for the starlight component.
In Sec.~\ref{sec:radiation-components}
we have already discussed the angle integrated spectrum 
in the solar neighborhood estimated by Mathis, Mezger and Panagia 
\cite{mezger1,Mathis:1983fx}. To model the
starlight spectrum for other points in the Galaxy we have used the results
of the same authors \cite{Mathis:1983fx} that give tables of the starlight spectra
for points on the Galactic equatorial plane ($z=0$) at different distances from 
the Galactic center, and the results on the starlight distribution
shown in the publications of MPS and ES.
The distribution of the starlight energy density in the Galaxy
has been described with the 
cylindrically symmetric and factorized form:
\begin{equation}
u_\lambda (r, z) = u_\lambda^{\rm GC} ~e^{-r/R -|z|/Z}~. 
\end{equation}
The parameters have been fitted as $R=2.17$~kpc (4.07~kpc) 
for $r < 8$~kpc, ($r > 8$~kpc) and $Z = 7.22$~kpc; the energy density at the
Galactic center is obtained imposing that the ISRF 
at the Sun position coincides with the estimate of Mathis, Mezger and Panagia.
For the angular distribution of the starlight photons we have assumed isotropy.
Our model for starlight in the Galaxy
should be considered as a simple first order approximation, but
it is adequate for the purpose of estimating the absorption of gamma rays, because
starlight represents only a very small correction to the contributions of the CMBR and
the infrared dust emission.
\begin{figure}[!t]
\includegraphics[width=8cm]{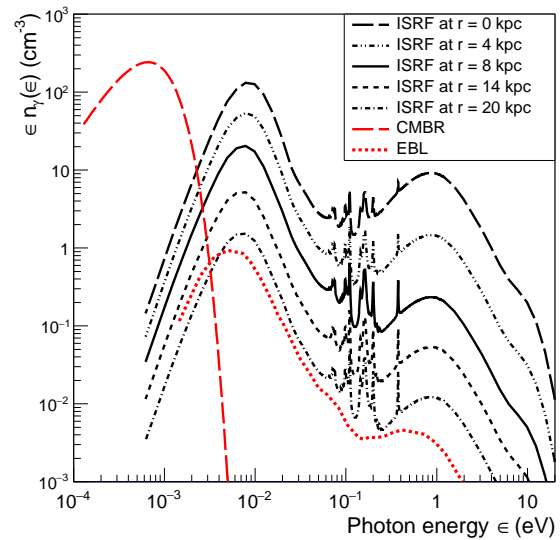}
\caption{Photon number density ($\varepsilon \, n_\gamma(\varepsilon)$ versus the
 energy $\varepsilon$) for points on the Galactic plane at different
 distances from the GC. The CMBR and EBL are shown separately
 and are independent from the position.
 The other curves represent the sum of the stellar and dust spectra.}
\label{fig:spectra}
\end{figure}

The energy spectra of the ISRF for points on the Galactic equator
at different distances from the Galactic center are shown in Fig.~\ref{fig:spectra}
plotted in the form $\varepsilon \; n_\gamma (\varepsilon)$ versus the
photon energy $\varepsilon$. Inspecting the figure one can see that the CMBR
is always the 
largest source of photons, followed by the infrared radiation emitted by dust,
with starlight contributing only a small fraction of the total number
of target photons.

\section{Absorption probability for Gamma Rays in the Galaxy}
\label{sec:absorption}

\begin{figure}[t]
\includegraphics[width=8cm]{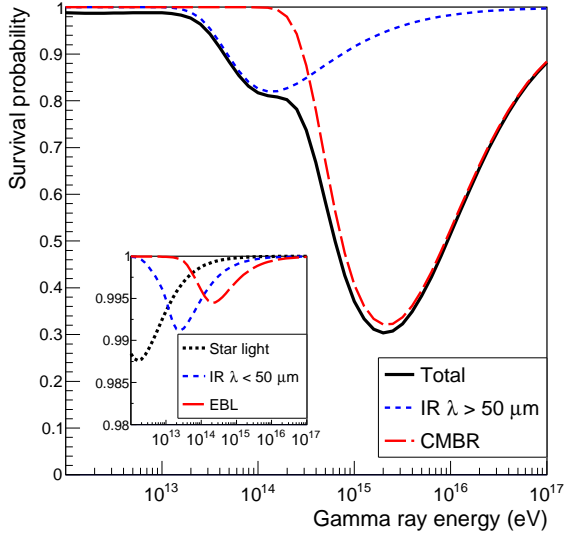}
\caption{Survival probability of gamma rays for a trajectory 
from the GC to the Sun, plotted as a function of the gamma ray
energy. The contributions of different radiation fields are shown. 
The inset shows the contributions of starlight, 
infrared radiation with wavelength $\lambda < 50~\mu$m and EBL.} 
\label{fig:abso_contr}
\end{figure}

Using the model for the ISRF described in the previous section, it
is straightforward to compute the survival probability
for gamma rays that travel different paths in the Galaxy
using Eq. \eqref{eq:prob-abs}.
Figure~\ref{fig:abso_contr} shows the survival probability
for gamma rays arriving to the Sun from a source in the Galactic center
as a function of the gamma ray energy.
In the figure the probability is shown together 
with the contributions from the different components of the radiation field.
Most of the absorption is due to the CMBR and to the thermal emission
from the dust (with wavelength $\lambda \gtrsim 50~\mu$m).
The other components of the ISRF give smaller contributions
that are visible in the inset of the figure.
The survival probability has a deeper minimum $P_{\rm surv} \simeq 0.30$ for 
 $E_{\gamma} \simeq 2.2$~PeV that is due to the CMBR, and a second minimum 
at $E_\gamma \simeq 150$~TeV. This structure can be understood
from the fact that the dominant sources of gamma ray absorption
are the CMBR and the infrared emission that have spectra
in different regions of $\varepsilon$.
The other components of the radiation give small corrections
to the absorption, indicating that the approximate treatments
of the starlight and line emissions from dust do not introduce
significant errors. The contribution of the
EBL is a correction of order $\Delta P \simeq 0.5$\%
for a gamma ray energy $E_\gamma \simeq 150$~TeV, where
its effects are most important.

Recently deep gamma ray observations by the
Cherenkov telescope H.E.S.S. \cite{HESS-GC}  in an annulus around
the Galactic center region have shown a spectrum that extends as a power law
up to energies of tens of TeV without indications of a break of cutoff,
strongly suggesting the existence of a proton ``PeVatron'' in the central 10~pc
of our Galaxy, probably associated to the supermassive black hole Sagittarius (Sgr) A*.
The study of this source with very high energy ($E_\gamma \gtrsim 30$~TeV) gamma rays
is clearly a crucial test, and a precise description of the absorption effects
is necessary. The study of the GC with neutrinos is also of great interest
(see for example \cite{Celli:2016uon}).

Gamma rays coming from different directions and distances 
have similar absorption patterns, with the maximum attenuation due to CMBR 
at $E_\gamma \simeq 2.2$~PeV, and a secondary absorption peak at
$E_\gamma \simeq 150$~TeV due to the
infrared light, that produces a ``shoulder'' in the total absorption spectrum. 
The amount of the two effects and their relative contributions
depend on the gamma ray path.

\begin{figure}[t]
\includegraphics[width=8cm]{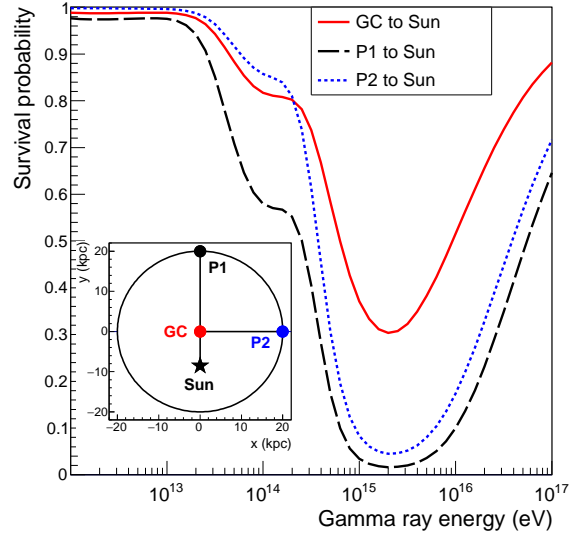}
\caption{Survival probability of gamma rays for three different trajectories 
in the Galactic plane, plotted 
as a function of the gamma ray energy. 
The inset shows the position of the sources.}
\label{fig:abso_3p}
\end{figure}

Figure~\ref{fig:abso_3p} shows the survival probability for three different source
positions: the Galactic center, the points P1 ($x=0$, $y=20$~kpc, $z=0$) and
P2 ($x=20$~kpc, $y=0$, $z=0$) (for the coordinate definition, see the inset of the
figure). 
The infrared absorption is maximum when the gamma rays 
arrive from P1, crossing the
Galactic center. Gamma rays arriving from P2 do not pass close
to the GC and the absorption effects due to infrared radiation are smaller.

\begin{figure}[t]
\includegraphics[width=8cm]{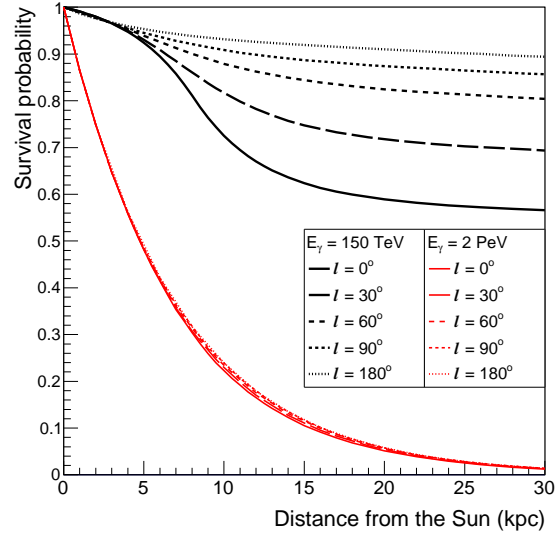}
\caption{Survival probabilities of gamma rays of energy 150~TeV and 2~PeV
 as a function of the source distance, for lines of sight with
 different Galactic longitudes and fixed latitude $b = 0^\circ$.}
\label{fig:abso_r_lat0}
\end{figure}

Figure~\ref{fig:abso_r_lat0} shows the gamma ray survival probability
for different directions in the Galactic equatorial plane, as a function of
the source distance, for gamma ray energies $E_{\gamma} = 150$~TeV and 2~PeV.
The absorption of gamma ray with $E_\gamma = 2$~PeV is mostly due to the
homogeneous CMBR, and is therefore to a good approximation
independent from the photon trajectory and described by a simple
exponential.
Gamma rays of 150~TeV are mostly absorbed by the infrared light,
and for a fixed source distance,
the absorption probability has a strong dependence on the
gamma ray path in the Galaxy, with the attenuation largest for
trajectories that cross the Galactic center.

\begin{figure}[t]
\includegraphics[width=8cm]{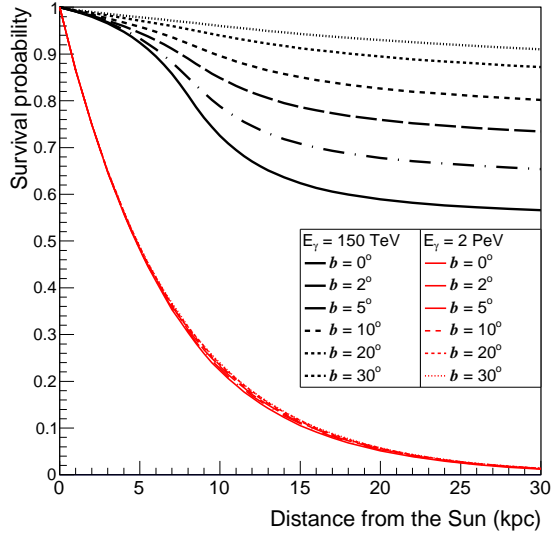}
\caption{Survival probabilities of gamma rays of energy~150 TeV and 2 PeV
 as a function of the source distance, for lines of sight with
 different Galactic latitudes and fixed longitude $l = 0^{\circ}$.}
\label{fig:abso_r_long0}
\end{figure}

The dependence of the absorption effect on the
direction of the gamma ray path, is also strong for lines
of sight that go out of the Galactic plane ($b \ne 0$),
as the density of infrared radiation is concentrated
near the Galactic plane.
The absorption of gamma rays for trajectories outside the Galactic plane
is illustrated in Fig.~\ref{fig:abso_r_long0} which shows the 
survival probability (for the same energies $E_{\gamma} = 150$~TeV and 2~PeV)
as function of distance for a set of lines of sight with different
Galactic latitude.

The gamma ray absorption probability $P_{\rm abs} (E_\gamma, b, \ell, d)$,
for a fixed value of the energy  $E_\gamma$, and a  fixed line of sight (determined
by the angles $b$ and $\ell$) grows  monotonically with the source distance  $d$.
Inspecting  Figs.~\ref{fig:abso_r_lat0}   and \ref{fig:abso_r_long0}   one can
see  that for large $d$ the probability associated to the Galactic radiation fields
becomes approximately constant, taking an asymptotic value that is a function
only of $E_\gamma$, $b$ and $\ell$. This is the consequence of the fact
that the density of Galactic photons falls rapidly ($\propto r^{-2}$)
at large distances from the center of the Galaxy,
and the absorption coefficient associated to Galactic photons
vanishes for $r \to \infty$.
The asymptotic value of the Galactic absorption probability
is appropriate for gamma ray sources located in the Galactic halo
at large distances from the Sun. This asymptotic  probability
is also necessary to  compute the total absorption of photons from
extragalactic sources combining this effect 
with the absorption in extragalactic space.

\begin{figure}[t]
\includegraphics[width=8cm]{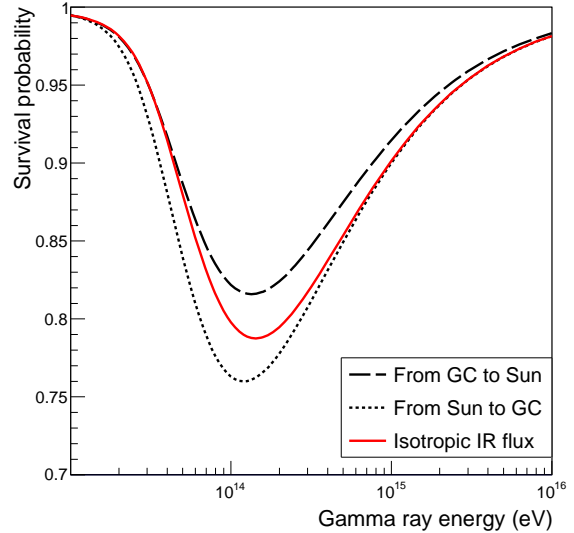}
\caption{Survival probability 
for gamma ray traveling from the Galactic center to the Sun, and from
the Sun to the Galactic center, as a function of the gamma ray energy,
calculated taking into account only the effects of dust emitted radiation.
The solid curve shows the survival probability
calculated assuming an isotropic distribution of the target photons.}
\label{fig:aniso}
\end{figure}

As already discussed, infrared radiation has a very strong anisotropy,
and the calculation of the survival probability has to include
the correct angular distribution of the target photons. 
To illustrate the importance of this effect,
Fig.~\ref{fig:aniso} shows the survival probability of a gamma ray
traveling from the Galaxy center to the Sun compared to that of
a gamma ray traveling in the opposite direction
taking into account only the contribution of the dust emitted radiation,
compared with a calculation that assumes
an isotropic angular distribution for the target photons.
As it is intuitive, the calculation with the assumption of isotropy
overestimates the absorption for gamma rays that travels
outward from the center to the periphery of the Galaxy
(when most of the target photons are parallel to the gamma ray direction)
and underestimates the absorption for gamma rays traveling in the
opposite direction. The effect is significant, but it represents
only a modest correction. For $E_\gamma \simeq 150$~TeV,
the absorption probability is $P_{\rm abs} \simeq 0.18$ for gamma rays traveling from the
GC to the Sun, and $P_{\rm abs} \simeq 0.24$ for gamma rays traveling in
the opposite direction, while the isotropic calculation is
approximately equal to the average.

\begin{figure}[t]
 \includegraphics[width=8cm]{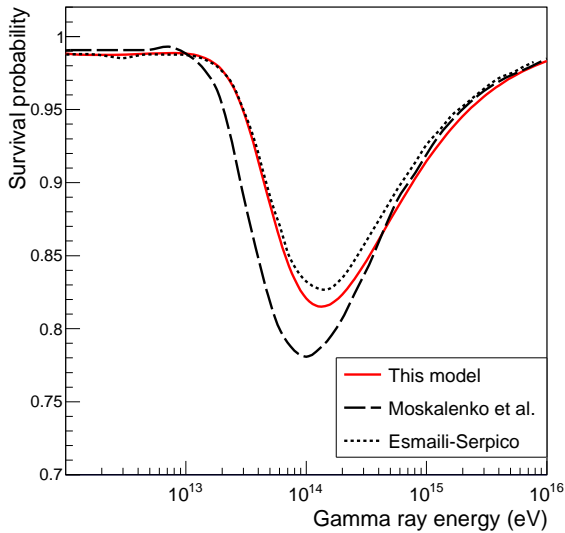}
 \caption{Survival probabilities for gamma rays traveling
 from the GC to the Sun (taking into account only dust and star emitted radiation).
 The different curves show our calculation and those of MPS
 \protect\cite{moska2006} and ES \protect\cite{esma2015}.}
\label{fig:abso_comparison}
\end{figure}

In Fig.~\ref{fig:abso_comparison} we compare the survival probability
calculated with our model for gamma rays that arrive from the Galactic center
with the results of MPS \cite{moska2006}
and ES \cite{esma2015}.
All calculations consider only the effects of
the radiation emitted by dust and stars.
The amounts of absorption are comparable in all three calculations, however some
differences can be seen. The calculation of MPS has a
larger absorption, which can be probably traced to the larger density
of infrared radiation estimated in the solar
neighborhood (see Fig.~\ref{fig:spectrum_here}).
The results of the ES calculation are similar to ours. 
This is likely due to a cancellation effect:
ES give a smaller estimate of the density of infrared radiation near the Sun,
but also assume that the radiation
is isotropic, overestimating the absorption
for photons that travel from the GC to the Sun.

\subsection{Systematic uncertainties}
\label{sec:systematic}
The crucial step in the construction of a model
for the infrared radiation in the Galaxy is the estimate of the
infrared emission density from the observations of
spectrum and angular distributions of infrared photons at the Earth.
This can be performed ``inverting'' Eq. (\ref{eq:flux-infra})
that gives the observed flux as an integral over the emission density.
Such an inversion is clearly a nontrivial problem.
Following Misiriotis et at. \cite{misi2006} we have used very simple and smooth 
(exponential) expressions for the space and temperature distributions of
interstellar dust, which describes reasonably well the main features of the data.
It is however clear that this simplified treatment has its limitations.

One source of error derives from the fact that the dust space distribution
is more irregular and has much more structure than a simple exponential.
The density of the radiation is smoother than the emission,
because each point in space receives photons from the entire Galaxy,
but some ``granularity'' in the density distribution of infrared radiation remains.
This has the consequence that the true optical depth along
a particular line of sight can differ from the
smoothed out estimate calculated in the model by as much as 20\%--30\%.

In addition it also possible that the exponential functional forms
that we have used do not capture exactly the average space and temperature
distributions of the interstellar dust, especially in the inner part of the Galaxy.
For example in \cite{Wright:1991zz}, the authors discuss a model for
the space distribution of the dust that has a (dominant) exponential
term together with a Gaussian, which enhances the dust mass density near the GC,
and also a narrow ``ring'' of dust of radius $r \simeq 0.50$~kpc.
Our model also neglects the spiral structures in the Galactic disk.

Numerical experiments performed along these lines suggest that
without spoiling (or in fact also improving) the agreement between data and model,
it is possible to change the optical depth for trajectories that traverse
the inner part of the Galaxy by something that can be as large as 30\%.

In our model the optical depth due to the infrared radiation
for a gamma ray source at the GC
has a maximum value of order $\tau \approx 0.2$ ($P_{\rm abs} \simeq 0.18$),
while the optical depth for a source on the
far side of the Galaxy is of order $\tau \approx 0.6$
($P_{\rm abs} \simeq 0.45$). A variation of $\pm 30$\% of the optical depth 
results for the first (second) case 
in a difference in the absorption probability 
$\Delta P_{\rm abs} \simeq \pm 0.04$ ($\Delta P_{\rm abs} \simeq \pm 0.08$).

\section {Outlook and Conclusions} 
\label{sec:outlook}
In this work we have calculated the attenuation of very high energy
gamma rays propagating in the Galactic and extragalactic radiation fields. 
The model for the Galactic infrared  radiation field
that is the crucial element in the  estimate of 
the absorption probabilities is based on a  simple  analytic description
of  the interstellar dust, and  has been described in a self--contained form.
It is therefore straightforward to replicate
the calculations performed here \footnote{To obtain tables
of the gamma ray absorption probabilities it is possible to contact the authors.}.

The CMBR is the most important source of target photons, and
with a spectrum that peaks at $\varepsilon \simeq 3.7 \times 10^{-4}$~eV,
is most effective for absorption at gamma ray energy $E_\gamma \simeq 2.2$~PeV,
where the absorption length is 7.07~kpc. This
implies that in the PeV energy range the Galactic center is visible 
with an optical depth of order unity, and galactic sources beyond
the Galactic center are very strongly attenuated.
At lower energy the effects of the CMBR are much less significant.
As an illustration, the absorption length at 300~TeV is 72~kpc, and the effects
decrease exponentially for decreasing $E_\gamma$.

For $E_\gamma \lesssim 300$~TeV, the absorption of gamma rays
is dominated by target photons generated inside the Galaxy. 
This radiation has two main components, starlight and dust emission,
which contribute an approximately equal energy density in all points of the Galaxy.
The spectra of the two components
are however well separated, with starlight photons having
an average energy of order $\langle \varepsilon \rangle \approx 1$~eV,
and dust emitted photons having an average energy 
$\langle \varepsilon \rangle \approx 2 \times 10^{-2}$~eV.
It follows that starlight photons are most effective
in the absorptions of gamma rays with energy $E_\gamma \sim 1$~TeV,
while infrared, dust emitted photons are mostly effective
for $E_\gamma \sim 150$~TeV.
The number densities of the target photons for the
two components differ by a factor of order 50, and therefore
the largest optical depths generated by the two components
differ by a factor of the same order.

The infrared radiation can generate optical depths as large as $\tau \approx 0.6$
(or absorption probabilities of order $P_{\rm abs} \approx 0.45$)
for lines of sight that traverse the entire Galaxy passing close to the center.
This is clearly an important effect, 
but the entire Galaxy and its halo can be observed also in the energy range
where the absorption due to infrared radiation is most important.
Absorption due to starlight remains always smaller
than 1\%--2\% for gamma rays created in any point of the Galaxy,
and is in practice of negligible importance.

In this paper we have discussed a simple model,
constructed following very closely the work
of Misioriotis et al. \cite{misi2006},
to compute the dust emitted radiation field using only few 
parameters. The model describes well the main features (in energy, space and angle)
of the radiation.
The original work of Misiriotis et al. considered only
the wavelength range $\lambda \gtrsim 50~\mu$m, where the dust emission
is well described as a thermal process with a 
modified blackbody spectral shape.
The emission at shorter wavelength is 
more difficult to describe, and more dependent on the (only poorly known)
composition of the dust grains.
Extending the work of Misiriotis et al. we have included in the model
a simplified description of dust emission
in the short wavelength range $\lambda \simeq 1$--50~$\mu$m 
that is however adequate for the purpose of computing
gamma ray absorption probabilities.

We have applied our model of the dust emitted radiation,
together with a simplified description of starlight,
and models of the CMBR and EBL to compute the absorption
probability for different gamma ray trajectories in the Galaxy.
Some examples of these calculations are shown in this work
and compared with results already existing in the literature.
The agreement is reasonably good.

A limitation of our model (that is also in common
with previous calculations) is that the interstellar dust spatial
distribution is modeled as a smooth exponential with cylindrical symmetry,
while the true distribution has a much more
complex and irregular form. It follows that our model
cannot claim to determine exactly the precise absorption for any
individual gamma ray source, however the results can be considered
valid for an ensemble of sources, or for the study
of a diffuse flux.

The main source of systematic uncertainty in the modeling
of the infrared radiation is probably the
description of the dust space distribution.
We have estimated the uncertainty for the 
optical depth due to infrared radiation
in the directions where it is largest
(when the line of sight passes close the GC)
as of order $\Delta \tau/\tau \approx 20\%$--30\%.
This corresponds to an uncertainty on the absorption
probability $\Delta P_{\rm abs} \lesssim 0.08$.

A comment is in order about the effects of the angular distribution
of the photons that form the ISRF.
The angular distribution of these photons is strongly anisotropic,
with the flux concentrated in a narrow band of Galactic
latitude, and a significant flow of the radiation outward from the Galactic center.
For gamma rays arriving from the Galactic center neglecting
the correct angular distributions of the target photons results in an error
for the absorption probability of $\Delta P_{\rm abs} \simeq 0.03$.

In conclusion, the absorption of very high energy gamma rays with
$E_\gamma \gtrsim 30$~TeV by pair production interactions effectively
precludes extragalactic observations,
but Galactic astronomy remains possible and very promising.
The pair production absorption effects can be calculated with an excellent
(for the CMBR), or reasonably good (for the dust emitted radiation) accuracy.

\end{document}